# A Conceptual Hybrid Framework for Post-Quantum Security: Integrating BB84 QKD, AES, and Bio-inspired Mechanisms

Md. Ismiel Hossen Abir

Department of Computer Science and Engineering

International Standard University

Dhaka, Bangladesh

Email: ismielabir286@gmail.com

## Abstract

Quantum computing is a significant risk to classical cryptographic, especially RSA, which depends on the difficulty of factoring large numbers. Classical factorization methods, such as Trial Division and Pollard's Rho, are inefficient for large keys, while Shor's quantum algorithm can break RSA efficiently in polynomial time. This research studies RSA's vulnerabilities under both classical and quantum attacks and designs a hybrid security framework to ensure data protection in the post-quantum era. The conceptual framework combines AES encryption for classical security, BB84 Quantum Key Distribution (QKD) for secure key exchange with eavesdropping detection, quantum state comparison for lightweight authentication, and a bio-inspired immune system for adaptive threat detection. RSA is vulnerable to Shor's algorithm, BB84 achieves full key agreement in ideal conditions, and it detects eavesdropping with high accuracy. The conceptual model includes both classical and quantum security methods, providing a scalable and adaptive solution for Post-Quantum encryption data protection. This work primarily proposes a conceptual framework. Detailed implementation, security proofs, and extensive experimental validation are considered future work.

# 1. INTRODUCTION

In 21$^{st}$ century, secure communication and information exchange is one of the most crucial to digital infrastructures worldwide. Everyday individuals, corporations, and government use cryptographic techniques to protect their sensitive data including financial transactions, and user authentication system. To protect sensitive information, classical cryptography algorithms such as RSA and AES (Advanced Encryption Standard) are used. These cryptographic systems use complex mathematical algorithms that are hard and sometimes impossible to break using classical computational resources. A key example is RSA algorithm which depends on the mathematical challenge of factoring large numbers [1]. The confidentiality, accurate, and authenticity of digital information depends on cryptographic mechanisms.

Standard cryptographic techniques, including RSA, AES remain secure in the classical computational settings. Despite this, the growth of quantum computing has created both new opportunities and significant challenges for classical cryptography. Quantum computer works differently compared to classical computers. Classical computers use bits that can be either 0 or 1. But quantum computers operate with qubits, that can be exist in the state of $|0\rangle$, $|1\rangle$, or $|\psi\rangle = \alpha|0\rangle + \beta|1\rangle$ superposition of both states [2]. Quantum computers can also use multiple qubits states, such as $|00\rangle$, $|01\rangle$, etc., therefore quantum computer can process many possible outcomes at the same time that helps calculation more efficiently than classical computers [2].

Quantum algorithms are capable of efficiently solving problems that are difficult for classical algorithms and are widely used in data security. One example is Sycamore quantum processor solved a problem in 200 seconds, where a supercomputer require nearly 10000 years, which shows the proof of quantum efficiency [3]. One significant quantum algorithms is the Shor algorithm, that demonstrate quantum computers can factor integers and solve discrete logarithm problems more efficiently than classical computers [4]. Shor algorithm can efficiently solve the mathematically problem that used in the RSA and others classical encryption systems [3]. For this reason, NIST recently finalized three post-quantum cryptography standards (PQC), such as FIPS 203, 204, and 205, to protect digital communications from quantum attacks [5]. Because the concerning part is if quantum computer is available, it could break the encryption protecting things such as online banking, secure communications, or sensitive government data. As a result, researches and teach companies are actively working on quantum advancements to ensure the future security of digital information. PQC approaches like code-based and lattice-based encryption to address data security challenges in the quantum era [6]. However, these approaches face challenges, such as code-based encryption has large key sizes and limited flexibility, while lattice-based encryption has less proven security and potential vulnerabilities [6].

## 1.1. Problem Statement

To overcome these challenges and enhance information security, we propose a hybrid model that involves Quantum Key Distribution (QKD) using BB84 protocol, Advanced Encryption Standard (AES), Quantum authentication mechansim, and bio-inspired mechanism based on immune system behavior. QKD such as BB84 protocol introduced by Bennett and Brassard in 1984 uses quantum mechanics to securely share encryption keys by using polarized photons [7]. This system provides security which is granted by physics law. From an encryption perspective, we more secure. This combination of BB84 and AES can give strong protection by using the strengths of both systems. We also include a quantum authentication component for identity verification. To make the system more secure, dynamic and intelligent we can use immune system mechanism which is inspired from biological system.

In this research, we explore the vulnerability of the RSA algorithm by analyzing both classical and quantum approaches. For the classical perspective we used trial division factors and Pollard Rho's factor, and the quantum perspective used the Shor algorithm. Then proposed a hybrid model is proposed that combines the classical AES encryption algorithm, the quantum BB84 protocol for key distribution, quantum authentication for identity verification, and a bio-inspired mechanism based on the immune system, to protect against attacks from both classical and quantum computers. The aim is to design this hybrid model to be resistant to both current and future cyber threats, particularly in the post-quantum era.

## 1.2. Objective of the Research

The advancement of quantum computing increases threats to classical cryptographic algorithms. In the post-quantum perspective, both classical and quantum security will face vulnerabilities and security issues. Therefore, post-quantum data security needs to implement a model that can handle classical, quantum, and post-quantum cyber-attacks and secure data. For this reason, primary goal is analyze the vulnerabilities of classical RSA algorithm using quantum Shor's algorithm. Then, design a hybrid cryptographic model that can ensure security against classical, quantum, and post-quantum cyber defense. The goal include in this research are

(i) Analysis vulnerabilities of classical RSA algorithm by using classical factorization methods and quantum Shor's algorithm.

(ii) To design a hybrid cryptographic model integrates Quantum Key Distribution (QKD) using BB84 protocol with Advanced Encryption Standard (AES) for enhanced data security and confidentiality.

(iii) Include a bio-inspired immune system mechanism for making the system adaptive and capable of detecting novel threats dynamically.

(iv) Using BB84, AES, quantum authentication mechanism, and bio-inspired immune system design a hybrid proposed model for ensure post-quantum security.

The proposed system aims to provide a secure, efficient, and post-quantum encryption framework that can address the challenges of the post-quantum era.

## 1.3. Overview of Challenges

Analysis of RSA vulnerabilities and implementation of BB84 protocols, we faced several challenges. Quantum computing is in its emerging stage, and simulating large-scale quantum algorithms requires significant computational resources. Current hardware limitations, such as noise, decoherence, and error rates, make practical implementation of quantum algorithms difficult. Furthermore, post-quantum cryptography standards are still evolving, which makes it challenging to design a system that will remain secure against future quantum advancements.

## 1.4. Contribution

The present work of this thesis structured into several key stages.

(i) RSA Vulnerability Analysis: Firstly, analysis RSA algorithm. Secondly, analyze its vulnerabilities using classical algorithms, including trial division factorization and Pollard Rho algorithm. Thirdly, analyze its vulnerabilities using the quantum Shor algorithm.

(ii) BB84 Quantum Key Distribution: Explain and implement secure key exchange using BB84 quantum protocol in simulated environment.

(iii) AES Encryption Using BB84: Explain the AES encryption algorithm and implement a secure communication system combining BB84 and AES.

(iv) Quantum authentication and Bio-inspired Security Layer: For second level security using quantum authentication for identity verification. After that, add an extra layer which is bio-inspired comes from immune system for dynamically monitors communication patterns, detects anomalies, and uses reinforcement to learn cyber threats and protect the system.

(v) Design of a Hybrid Model: Using BB84, AES, quantum authentication, and bio-inspired immune system to design a hybrid model to ensure data security among the classical, quantum, and post-quantum eras.

## 1.5. Fundamentals

### 1.5.1. Classical Cryptography

In the classical cryptography perspective, use the RSA and AES algorithms. Classical cryptography mainly used mathematical complexity computation such as factorize large number to secure communication [8]. In this research from the classical cryptography part used RSA (Rivest-Shamir-Adleman) and AES algorithm. RSA uses asymmetric cryptography, where both public key and private key applied for factoring large prime numbers [8]. AES is a symmetric key encryption that used AES-128, AES-192, etc. fixed size blocks for encrypt data [9].

### 1.5.2. Quantum Computing
Quantum computing follows different mechanism than classical computer. Instead of using bits quantum computer used qubits, where used a superposition state and quantum computing also implement quantum gate such as Hadamard gate, Pauli-X gate, Controlled-NOT gate, etc. [10]. This properties increase the computational and process many possibilities simultaneously. Quantum algorithms such as Shor's factorization can break classical RSA algorithm.

### 1.5.3. Quantum Key Distribution BB84 Protocol
Quantum Key Distribution (QKD) create secure encryption keys which is significantly different from classical methods [11]. One of the popular QKD method is BB84, which proposed by Bennett and Brassard in 1984, that uses polarization states photon for encode information and detect any eavesdropping attempt [12]. The BB84 protocol follows no-cloning theorem that ensures eavesdropping attempt can be identified [13].

### 1.5.4. Quantum Authentication Mechanism
We incorporate a quantum authentication mechanism in our framework that utilizes quantum states for identity verification. This approach is conceptually inspired by quantum communication protocols that offer efficiency advantages. In this approach inspired from quantum fingerprint. Quantum fingerprint demonstrate that, it is exponentially shorter than classical fingerprint which uses multiple channels to send information faster [14]. This method can reduce communication time and needs lower photons that increase the efficiency compared with classical method [14]. In our authentication mechanism, we use similar quantum-state principles to create verification tokens that can be efficiently compared for authentication purposes.

### 1.5.5. Bio-inspired Security Mechanism
Based on the immune system mechanism, it provides inspiration for developing dynamic and intelligent security mechanism. In this research, uses human immune system pathogen recognition capacity in our hybrid model to add an extra layer of security. Here, our model will recognize, respond to, and remember any threats and harmful activities. The system will uses reinforcement learning, anomaly detection, and distributed defense mechanisms.

## 1.6. Recent Works' Review
Several existing recent work and original works related to this research explored including RSA, AES, Shor's algorithm, BB84 protocol, quantum fingerprint, and immune system mechanism to combine a hybrid model to secure information.

Several recent research works explores the quantum computing threats on classical computing. Quantum computer can break classical cryptography that is a threat for secure information but QKD helps to secure the information [15]. Chen et al., study also mention that, China developed the China Quantum Communication Network for secure network that uses QKD with satellites and fiber to protect messages across 10,000 km, 17 provinces, and 80 cities [15]. Quantum computing is also threat to healthcare data; therefore, Poorani and Anitha's research suggests a new system using strong encryption, blockchain, and special key generation to protect medical information [16]. Therefore, to protect data from every sector, we need to use new model that can secure our information and can survive in the PQC. Shor factorizing algorithm can solve the mathematically problem of RSA algorithm [3]. Kim et al., in their research develop cloud file-encryption system using PQC, BB84 from QKD, and AES that speeds up encryption up to 8.11 times and overall runtime by 2.37 times [17]. This research explores post quantum encryption to secure information. Some research explore quantum fingerprint such as Ziiatdinov et al. developed quantum fingerprinting method that using additive combinatorics to optimize circuit depth, and experiments on IBMQ simulators that works better with fewer errors [18]. A recent study proposed secure method for MANET routing that uses trust, energy saving, digital signatures, and encryption to improve security [19].

The recent research focuses on individual solutions such as some can solve classical problem but quantum approach can break it. On the other hand, some research focus on individual either QKD or PQC. But in this research, we combines quantum and classical security methods with bio-inspired features to create hybrid systems that protect against current and future threats. This research aims to fill the gap by creating a hybrid model that combines BB84 quantum key distribution, AES encryption, quantum authentication, and bio-inspired immune system mechanisms to provide complete protection against classical, quantum, and post-quantum attacks.

## 2. LITERATURE REVIEW

This chapter presents a thorough review of existing research related to quantum computing, classical cryptography and vulnerabilities on quantum algorithm, and security models. The literature review in this section will cover the main topics of this research, including RSA algorithm vulnerabilities, Shor's factorization quantum algorithm, post-quantum cryptography solutions, quantum key distribution methods, quantum fingerprinting techniques, and bio-inspired security mechanisms. The review helps to understand the current state of research and identifies gaps that this study aims to fill.

## 2.1. Classical Cryptography

Taherdoost et al., reviews 495 studies based on the use of cryptography to enhance AI security and they found rapid research growth in this field from 2020 such as cryptography, blockchain, AI security, etc. [20]. Another recent study focus that, due to the increase of internet use for communication to transactions, we need to improve data confidentiality and integrity by using symmetric cryptography, asymmetric cryptography, RSA, Diffie–Hellman, DSA, etc. [21]. Another paper explore that cryptography technique is essential for both wired and wireless network against cyber security by using RSA, AES, etc. algorithm [22]. One recent study focus on cryptography essentiality in cloud computing by ensuring authentication, confidentiality, integrity, and availability, with methods like DNA cryptography, ECC, etc. methods [23]. Another paper explore about cognitive cryptography technique where they showed that secures strategic information by splitting it into parts that distributed among trustees, using biometric traits and semantic features [24]. Cryptographic techniques protect information by ensuring confidentiality, integrity, and authentication, with key management being critical to security [25]. In this research Zhou and Tang practically implemented RSA algorithm [25]. Gupta et al., in their research reviews and categorizes existing image cryptography techniques using a new taxonomy based on confidentiality, integrity, and authenticity [26]. They analysis performance and security through evaluation protocols and cryptanalysis methods, explores deep learning-based approaches [26].

Chang et al., in their research, they improve the AES and RSA algorithms to make it faster and energy-efficient for IoT devices by using a three threaded AES and a triple prime RSA design [27]. In this research, they mainly combined the RSA-AES algorithm, making an improved hybrid algorithm for securely encrypting data [27]. Jiang et al., in their research introduces MFDRL-RSA, a deep reinforcement learning method using multi-link fragmentation to improve routing [28]. Another work explore RSA where they tried to develop a secure communication by using graph theory, RSA encryption, and triangular fuzzy membership functions into a fuzzy graph network [29]. Somsuk proposes improved equations for RSA digital signatures by using a smaller integer instead of the private key in signing and they achieving 30% faster efficiency using the proposes method [30]. Helmy explore a hybrid audio encryption method combining AES and plexus algorithms and found that encryption improve strength and transmission quality [31]. Huo et al., in their research introduces a simplified AES image encryption method using a three-dimensional hyperchaotic system and they achieving up to 87% faster encryption and strong resistance to cryptographic attacks compared to traditional AES [32]. But due to the advancement of quantum computing, classical cryptography may face difficulties in ensuring security. We observe that Shor's algorithm can break several classical cryptography algorithms.

## 2.2. Quantum Computing

Quantum computing follows different approach computation compared to classical computers. Recent study of quantum computing discusses quantum algorithms including Shor's and Grover's, and quantum cryptography, such as QKD and post-quantum cryptography for secure communications [33]. They also highlights quantum hardware challenges including, decoherence, error correction, and future applications in AI, climate modeling, and cyber security [33]. Another research, explore the quantum computing benchmark methods to measure the performance of quantum computers, focusing on error rates, speed, and scalability [34]. They also compare between high and low level benchmark and predict a quantum computer's real-world usefulness [34]. Some study explores distributed quantum computing. Barral et al. explores the current state of distributed quantum computing, and computational power where they focus on four layers such as physical, network, development, and application to explain key concepts, challenges, and advancements in the field [35]. Some research works explore Quantum Machine Learning (QML), quantum chemistry and quantum advancement algorithms. Patel et al. reviews modern measurement techniques for quantum chemistry, especially for the Variational Quantum Eigensolver, explaining methods to reduce cost, improve accuracy, and handle errors [36]. Cerezo et al. explored QML methods and applications combined with quantum computing and Machine Learning (ML) they tried to speed up data analysis [37]. Martín-Guerrero & Lamata explored quantum computing and ML techniques that solve complex problem [38]. In this study, they mainly focus on two main approaches such as ML in quantum computing perspective and applying quantum computing to improve classical computing tasks [38]. Some research explores quantum neural network perspective. Wiebe et al. proposed two quantum algorithms that reducing the computational and statistical complexity [39]. In the first algorithm, they explored separating hyperplane and found (O√N) quantum amplitude amplification and in the second algorithm, they tried to improve classical issues by using bound from $(1/\gamma^2)$ $O(1/\sqrt{\gamma})$ where $\gamma$ refers as margin [39]. Sagingalieva et al. proposed hybrid QNN model that can predicts cancer drug responses better than classical models and they found 15% improvement than classical models using less training data [40]. In this work for drug response prediction, they used deep quantum neural with 363 layers and 8 qubits circuits [40].

Quantum computing and its algorithms are essential for secure information. Subramani et al. in their research, compares classical and quantum cryptography by reviewing their methods, security levels, and applications, analyzing various protocols, and giving recommendations for choosing the best encryption model for secure communication [41]. Several recent researches work explored and showed that Shor's algorithm can break RSA algorithm. Liu et al. improves Shor's algorithm for breaking RSA by designing quantum circuits with fewer CNOT gates [42]. In this research, they also analyzed running time on ion-trap quantum computers, and estimates the feasibility of factoring large integers more efficiently [42]. Albuainain et al. in their research explores quantum computing concepts and demonstrates how Shor's algorithm can break RSA encryption faster than classical methods [43]. A recent research explores Shor's algorithm implementation where they used quantum phase estimation and period-finding techniques on a simulated 6-qubit quantum computer to factorize the number 35 [44]. They also highlights the risk of quantum computer and suggest post-quantum cryptography to secure data in the future [44]. Willsch et al. in their research used large-scale GPU-based simulator and testing over 60,000 factoring problems by using Shor's quantum factoring algorithm [45].

## 2.3. Quantum Key Distribution

Quantum Key Distribution (QKD) used different types of approach aim to secure information in quantum era. Liu et al. in their research explore twin-field QKD and experiment using over 1002 km of optical fiber using 3-intensity protocol and advanced phase stabilization [46]. Their implementation achieved a secure key rate of $9.53 \times 10^{-12}$ per pulse at 1002 km and 47 kbps over 202 km, proving quantum-secure communication [46]. Liao et al. in their research demonstrated satellite-to-ground QKD over 1200 km using China's Micius satellite, and they achieved secure key exchange with rates up to 1.1 kbit/s [47]. They used the BB84 protocol with polarization encoding to maintain a stable quantum [47]. Yuan et al. used transmit quantum information using the BB84 protocol [48]. They also showed stable and secure way to send quantum information wirelessly [48]. A research used QKD protocols such as BB84, BB92, etc. to alternative of classical cryptography [49]. Another research explore combines the BB84 protocol with AES encryption to create a highly secure method [50].

## 2.4. Quantum Authentication and Bio-inspired Security

Quantum fingerprint can increase authentication and secure information. Ziiatdinov et al. in their research, they explored a method for quantum fingerprinting that uses shallow circuits and their method efficient than the current method [18]. Another research showed that quantum fingerprints exponentially smaller than the original strings [51]. A research shows that quantum fingerprints can solve a wide range of functions using only O(log n) bits that is fewer than classical methods [52]. Biological immune system can be efficient method for secure information. There are several bio-inspired encryption mechanism. He et al. in their research developed $Zn_{1.2}Ga_{1.6}Ge_{0.2}O_4:Ni^{2+}$ nanoparticles based advanced optical information encryption and achieved improved security in dynamic environments [53]. Basu et al. proposes a bio-inspired cryptosystem that uses DNA, RNA, and protein coding processes for encryption and decryption, and they achieved efficient data security [54].

## 2.5. Research Gap and Limitations

Although many studies followed classical cryptographic algorithms such as RSA, AES, recent developments of quantum computing nowadays many studies follow QKD. Some research follow combine model using QKD and AES algorithm to secure information. But these methods have some limitations and concerns regarding whether it secure information in Post Quantum Cryptography. Therefore, to make data secure and survive in (PQC) in this research developed a combined quantum cryptography technique using BB84, AES, quantum authentication, and bio-inspired immune system method to ensure adaptive security against current and future threats.

## 3. CLASSICAL AND QUANTUM CRYPTOGRAPHY

The backgrounds cover RSA, AES cryptographic algorithms, trail division and Pollard Rho's factors, Shor's algorithm, QKD BB84 protocol, quantum authentication, and immune system concepts.

### 3.1. RSA Algorithm

The RSA name comes from three scientists, Rivest, Shamir, and Adleman, who developed this algorithm, which is asymmetric cryptographic algorithms [1]. RSA algorithm used large prime number, and modular arithmetic [1]. This algorithm depends on complex of factoring the product of two large prime number such as p and q and compute as, $n = p \times q$. Here, n is modulus and used for both encryption and decryption operation. RSA algorithm mainly works with number so its turned characters into ASCII numbers then operation. Then, RSA algorithm used Euler's totient function as φ(n):

$$\varphi(n) = (p-1)(q-1)$$

After that, the public key is composed as (e, n) where e is exponent and e should be:

$$1 < e < \varphi(n) \text{ and}$$
$$\gcd(e, \varphi(n)) = 1$$

Similarly, the private key is composed as (d, n) where d is the inverse of e with respect to φ(n). So d:

$$d = e^{-1} \pmod{\varphi(n)}$$

So, the encryption of plain text message is M and its cipher text is C as decryption methods of the RSA algorithm:

$$\text{Encryption, } C = M^e \pmod{n}$$
$$\text{Decryption, } M = C^d \pmod{n}$$

The security depends on RSA algorithm factorization problem. Therefore, it used larger prime number and after the product a larger modulus where it can be more than 2048 bits.

### 3.2. Advanced Encryption Standard (AES)

AES is a symmetric key encryption which used block size 128 bits and data length can be 128, 192, or 256 bits [56]. In this algorithm also used encryption rounds which is depends on key size such as for 128 bits 10 rounds, for 192 bits 12 rounds, etc. [9]. AES algorithm used several types of operations such as changing each byte using a substitution table, shifting the rows, mixing the columns, and adding a round key with XOR [9]. This algorithm is secure against classical encryption system and widely used for secure communication, file encryption.

### 3.3. Classical Factorization Method

In this research from the classical factorization perspective used two different methods such as Trial Division and Pollard's Rho Algorithm. Firstly, trial division is the basic level of factorization where used a number n and divisibility by using odd integer less than or equal to $\sqrt{n}$ [55]. The time complexity of this method is $O(\sqrt{n})$ and the process is slow for large number. Pollard's Rho is another factorization algorithm which is faster and efficient method than trial division because Pollard's Rho used n in $n^{1/4}$ [56]. But the improved version of this algorithm is approximately 24% faster [56]. This method used polynomial function for generate sequence of numbers and used Floyd's cycle detection algorithm for detecting cycles of sequences. After that, the algorithm used GCD operation to extract a non-trivial factor.

### 3.4. Quantum Computing

Quantum computing follows different approach than classical computing. Quantum computing used qubits which can be $|0\rangle$, $|1\rangle$ or superposition state [2].

$$|0\rangle = \begin{bmatrix}1\\0\end{bmatrix} \text{ and } |1\rangle = \begin{bmatrix}0\\1\end{bmatrix}$$

It also can be superposition by using both of the states:

$$|\psi\rangle = \alpha|0\rangle + \beta|1\rangle$$

Here, α and β is the complex number and it's normalization condition:

$$|\alpha|^2 + |\beta|^2 = 1$$

Here, $|\alpha|^2$ is the probability of measure of state $|0\rangle$ and $|\beta|^2$ is the probability of measure of state $|1\rangle$ [2].

We can use it for multiple qubits. For example: if we used two qubits:

$$|\psi\rangle = \alpha_{00}|00\rangle + \alpha_{01}|01\rangle + \alpha_{10}|10\rangle + \alpha_{11}|11\rangle$$

Also, its normalization is:

$$|\alpha_{00}|^2 + |\alpha_{01}|^2 + |\alpha_{10}|^2 + |\alpha_{11}|^2 = 1$$

For every n qubits, quantum computer can represent as $2^n$ states, which grows exponentially, and this is the reason it is faster.

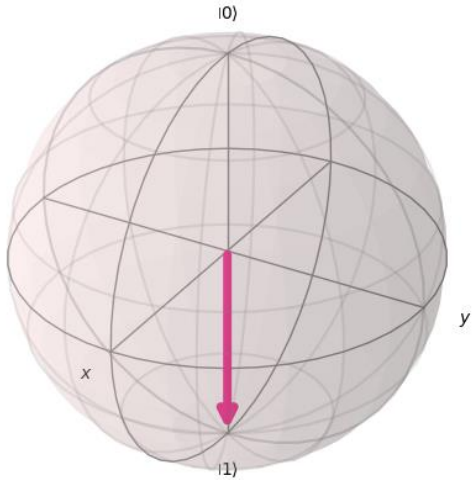

Figure 3.1. Pauli-X Gate

Quantum computing uses quantum gates that manipulate qubit states. There are several types of quantum gate such as Hadamard gate, CNOT gate, Rotation gate, etc. One of the basis and single qubit quantum gate is Pauli-X gate which flip the state of qubit such as $|0\rangle$ flip into $|1\rangle$ and $|1\rangle$ flip into $|0\rangle$. Some Pauli Gates are:

$$\text{Pauli-X gate} = \begin{bmatrix} 0 & 1 \\ 1 & 0 \end{bmatrix}, \quad \text{Pauli-Y gate} = \begin{bmatrix} 0 & -i \\ i & 0 \end{bmatrix}, \quad \text{Pauli-Z gate} = \begin{bmatrix} 1 & 0 \\ 0 & -1 \end{bmatrix}$$

In the Figure 3.1, Pauli-X gate shows Bloch sphere where input 0 and it showed output as 1. One of the most used gates is Hadamard gate, which used for creates superposition:

$$\text{Hadamard gate} = \frac{1}{\sqrt{2}} \begin{bmatrix} 1 & -1 \\ 1 & 1 \end{bmatrix}$$

Where,

$$H|0\rangle = \frac{|0\rangle + |1\rangle}{\sqrt{2}} \text{ and } H|1\rangle = \frac{|0\rangle - |1\rangle}{\sqrt{2}}$$

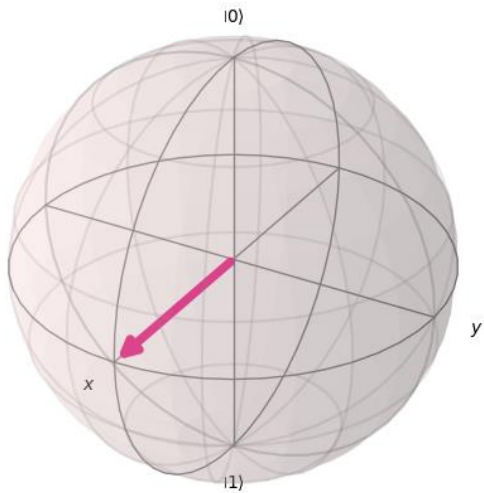

Figure 3.2. Hadamard Gate

Figure 3.2 represent hadamard gate Bloch sphere where input was 0 and the output shows superposition.

Another mostly used quantum gate is CNOT that used 2 qubits. CNOT gate matrix:

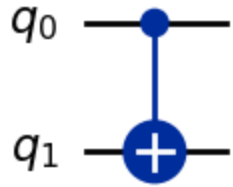

Figure 3.3. CNOT Gate

Figure 3.3 represent CNOT gate where two qubit gate that flips the target qubit if the control qubit is $|1\rangle$.

Quantum entanglement is a phenomenon where qubits become correlated and each of the qubits depends on each other. One of the most common entanglement example is Bell state where used Hadamard gate in the first qubit and CNOT gate on two qubit.

$$|\Phi+\rangle = \frac{|00\rangle + |11\rangle}{\sqrt{2}}$$

This means if the first qubit is measured as $|0\rangle$, the second will also be $|0\rangle$. If the first is $|1\rangle$, the second will be $|1\rangle$. Figure 3.4, represent the Bell state entanglement:

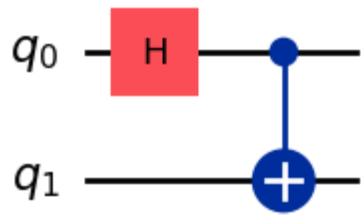

Figure 3.4. Bell State Entanglement

## 3.5. Shor's Algorithm

Shor's algorithm is the most crucial quantum algorithms because it can factorize very large integers much faster than any known classical algorithm [57]. For this reason, it can break RSA encryption method and concern for classical cryptography system. In this algorithm used two main components such as classical preprocessing and quantum subroutine. In this algorithm, firstly chose an integer a < N and then find period r of function [58].

$$f(x) = a^x \bmod N$$

After estimating r, a factor of N is computed using GCD:

$$\gcd(ar/2 \pm 1, N)$$

The quantum part of the algorithm efficiently finds the period r using the Quantum Phase Estimation (QPE), which includes modular exponentiation followed by the Quantum Fourier Transform (QFT). This step transforms the state of the qubits into a superposition that encodes information about the period, which can then be extracted with high probability after measurement.

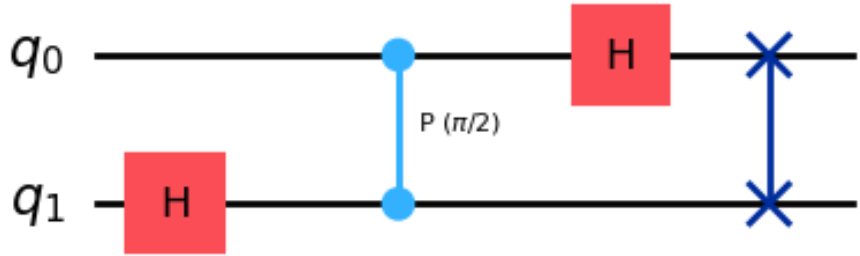

Figure 3.5. QFT Circuit for 2 Qubits

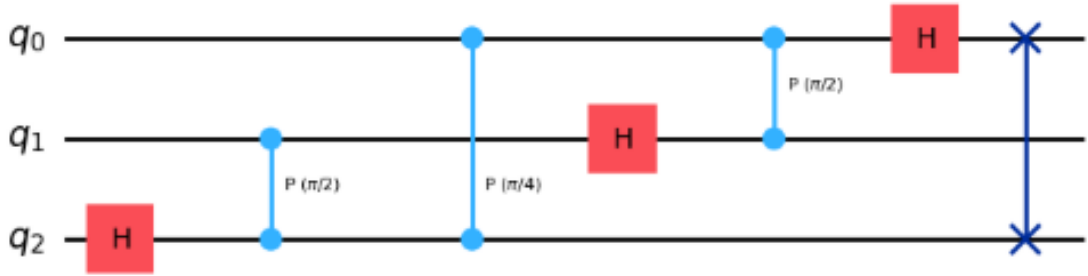

Figure 3.6. QFT Circuit for 3 Qubits

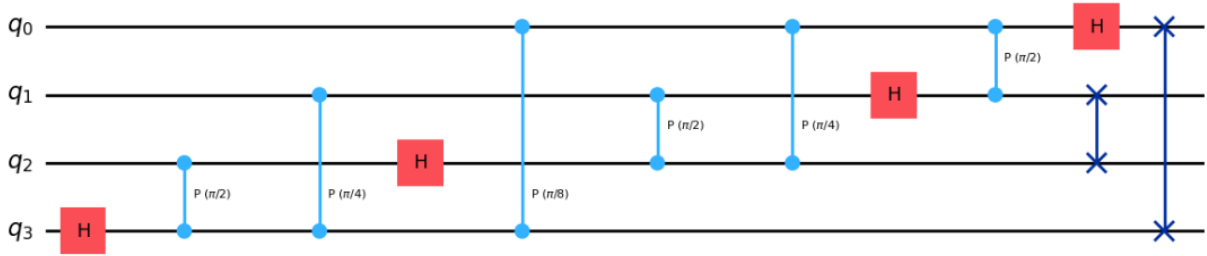

Figure 3.7. QFT Circuit for 4 Qubits

Here, Figure 3.5, 3.6, 3.7 illustrate QFT circuits for different numbers of qubits. Each circuit consists of three main steps such as hadamard gate, Controlled-phase rotations P(θ) for encode relative phase information between qubits, and swap gate. This algorithm runs in polynomial time $O((\log n)^3)$, making it exponentially faster for large numbers.

### 3.6. Quantum Key Distribution: BB84 Protocol

BB84 protocol is one of the most studied QKD. In BB84 protocol which distributed secrete key between Alice and Bob [59]. In this protocol if Eve tried to interact the communication it can be also detectable. This protocol utilized the no-cloning theorem and the uncertainty principle. In this protocol firstly, Alice randomly generates keys and bases which can be rectilinear (+) or diagonal (×) [59]. Then, Alice encodes keys using photon polarization states. After that, Alice sends it to bob using quantum channel and bob randomly measures the keys for decode. Alice and Bob compare their bases over a public channel and keep only the bits where the bases match. If the keys not match that indicates possible eavesdropping by Eve.

## 3.7. Quantum Authentication and Biological Immune System Mechanism

Our quantum authentication approach is inspired by quantum fingerprinting and uses this mechanism, and includes others. Quantum fingerprint mainly make larger dataset in a small fingerprint type for which is exponentially smaller than its original string [51]. Each dataset is converted into a compact quantum state, and fingerprints are sent through quantum channels. This approach reduces the amount of information needed and efficiently handles large amounts of data.

Another security mechanism in this research, we will use which is come from bio-inspired immune system. It will be adaptive and dynamic security system to protect data. This system will be identifying threats and response against threats. Then, the system memorized this type of threats. Therefore, in future the similar types of threats it can secure it efficiently. The system will also adaptive learn and evolve to handle new threats as like reinforcement system.

## 4. METHODOLOGY

In this methodology section, we explore this research step by step. We divided this methodology section into four sub-sections. In the first subsection, we explain this research and design it. Then, we will analysis the RSA algorithm vulnerabilities both classical and quantum computing algorithm perspectives. After that, explain and implement BB84 protocol. And in the last subsection, we will explore the combined model using AES, BB84 Protocol with Quantum authentication and Bio-Inspired Security.

This research follows an experimental design approach where we explore different security systems to build a comprehensive hybrid model. In this research we implement some methods and theoretically analysis some methods. According to the Figure 4.1 workflow diagram it describe about our methodology and results part. The research methodology is structured into four sequential phases: RSA vulnerability analysis, BB84 quantum key distribution implementation, combined BB84 protocol with AES encryption, quantum authentication, and bio-inspired security development. We used this step by step, because each phase builds on the previous one. First, we explore how RSA can be broken. Then we implement quantum key distribution to create secure keys. Next, we use these quantum keys with AES encryption to protect data. We add quantum authentication and bio-inspired security for extra protection. Finally, we combine everything into one complete system for secure information in the PQC era. For the simulation and analysis purpose, we used Google Colab for simulation, Qiskit library, Cirq library, Visual Studio Code. In this research, we used mixed method because each security component requires different testing methods and our main aim is to make it secure in the PQC era.

## 4.1. Research Design and Approach

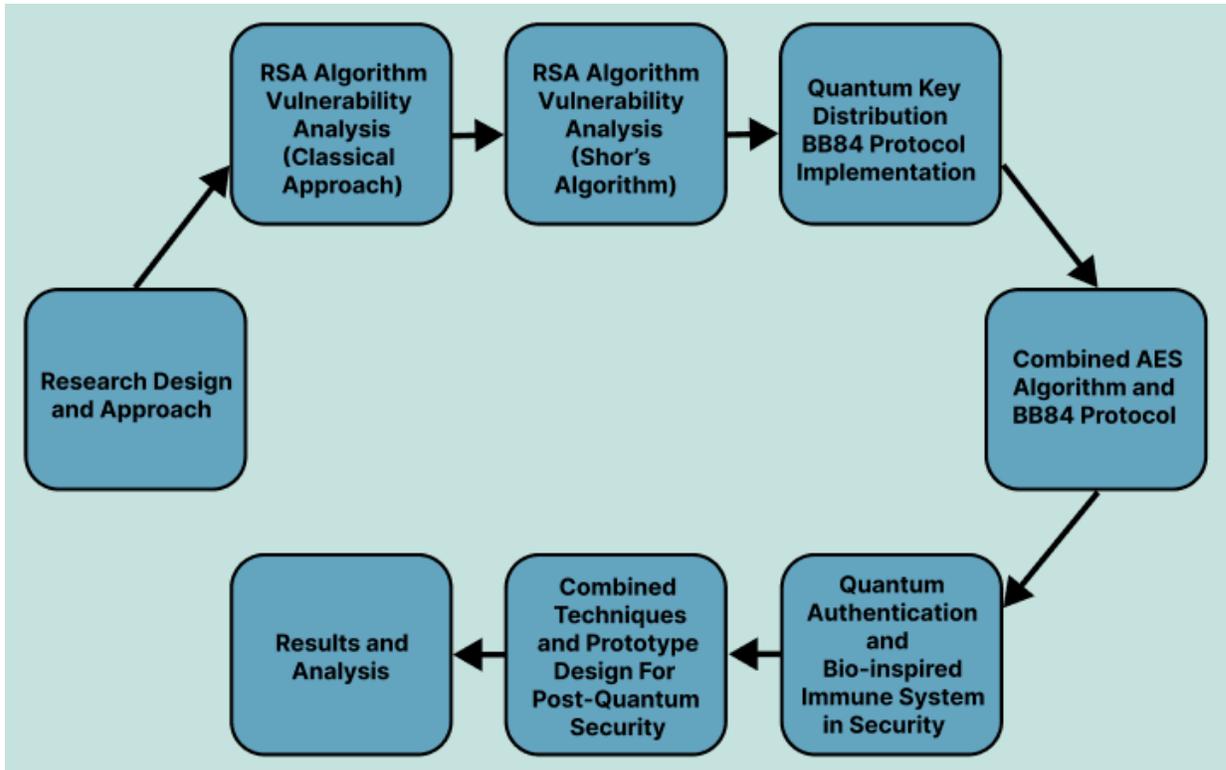

Figure 4.1. Workflow Diagram

## 4.2. RSA Vulnerability Analysis Classical and Quantum Perspective

In this phase, we first implemented the RSA algorithm and then tested classical factorization algorithms to demonstrate RSA vulnerabilities against classical attacks. We focused on two main methods: trial division factorization and Pollard's Rho. After that, we used Shor's quantum algorithm to explore the RSA algorithm vulnerability analysis.

### 4.2.1. RSA Key Generation

In this stage, we generated RSA key pairs using selected 20 prime number pairs to cover a wide range of RSA modulus values $N = p \times q$. Aim was to identify the correlation between key size and probability of classical and quantum attacks. To select prime number pairs, we divided them into four categories such as small prime numbers <100, medium prime numbers 500-1000, large prime numbers 50000-100000, and very large prime numbers >500000. Then, implemented RSA algorithm.

### 4.2.2. RSA Algorithm Step

Step 1: Select two prime numbers p and q according to the desired security level.

Step 2: Compute the RSA modulus: N = p x q.

Step 3: Compute Euler's totient: φ (N) = (p-1) x (q-1)

Step 4: Choose a public exponent as e = 65537 which is commonly used value for efficiency and security

Step 5: Compute the private exponent as d = modular inverse of e module φ (N) such as
(e x d) = 1 mod φ (N)

Step 6: Return the public key (N, e) and private key d.

### 4.2.3. RSA Encryption and Decryption Process

For each RSA key pair generated, we used the encryption and decryption process using plaintext message M = 42, e and N public key parameters and d is the private key.

The encryption process, $C = M^e \mod N$

The decryption process, $M_{decrypted} = C^d \mod N$

This verified the correctness of each key pair, demonstrating that RSA operates as expected for all selected key sizes.

### 4.2.4. Classical Factorization Analysis

In this part, we used two classical factorization algorithms Trial Division and Pollard's Rho, to check the classical factorization attacks of our RSA module.

**Algorithm 1: Trial Division**

Step 1: Set limit = floor(√N) + 1

Step 2: For each integer i from 2 to limit:
   If N mod i == 0:
     Return factor i

Step 3: If no factor found, return None

Trial division is simplest technique for factoring integers. It sequentially checks whether N is divisible by any integer between 2 and √N. The algorithm halts if it finds the first divisor. Time complexity is O(√N).

**Algorithm 2: Pollard's Rho**

Step 1: If N mod 2 == 0:
   Return 2

Step 2: Randomly select:
   x => random(2, N-1)
   y => x
   c => random(1, N-1)

Step 3: Define f(x) = (x^2 + c) mod N
Step 4: Repeat:
   x => f(x)
   y => f(f(y))
   d => gcd(|x - y|, N)
   If d == N:
     Return None
   If d > 1:
     Return d

Pollard's rho is a probabilistic algorithm that searches for factors using a pseudo-random sequence. It detects factors when two sequence values collide under modular arithmetic, causing the GCD of their difference with N to reveal a divisor. Time complexity is approximately $O(N^{0.25})$.

**4.2.5. Shor's Factorization Algorithm**

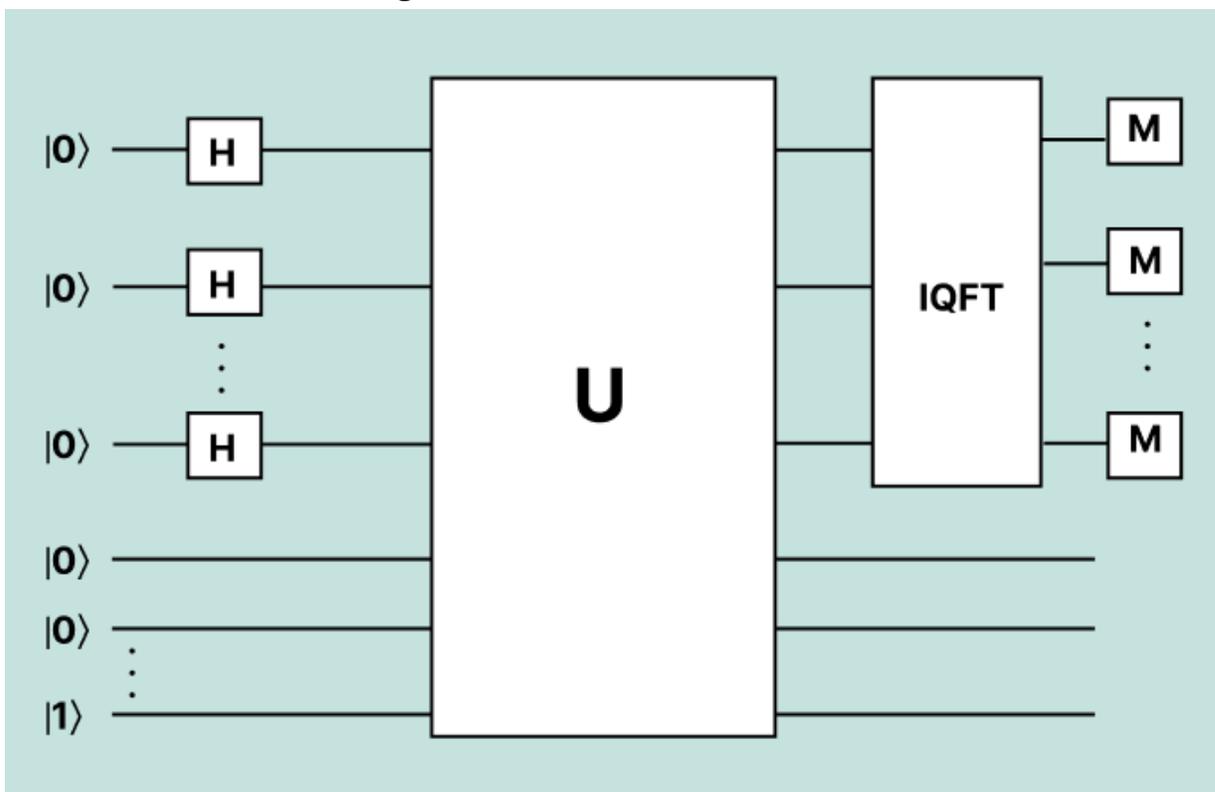

Figure 4.2. Shor Algorithm

After the check of classical factorization algorithm, we found that RSA algorithm can survive in the classical attack. Then, we used quantum Shor's algorithm to check the RSA algorithm vulnerabilities. The Figure 4.2 represents the Shor's algorithm where used two register first three source and others target register. In the source register, Hadamard gates are applied to create an equal superposition of states. Then, a unitary operation U is applied, which represents modular exponentiation. Unitary operation is the key step for finding the periodicity required for factorization. After this step, the inverse Quantum Fourier Transform (IQFT) is applied to the source register to extract the period information. This process helps in efficiently finding the factors of a large integer that is the main goal of Shor's algorithm.

**Steps of Shor's Quantum Algorithm**

Step 1: Classical preprocessing:
    Select random integer a where $1 < a < N$
    Compute $\gcd(a, N)$
    If $\gcd(a, N) > 1$: Return $\gcd(a, N)$

Step 2: Quantum period finding:
    Initialize quantum registers $|x\rangle|y\rangle$
    Create superposition: $H^{\otimes n} |0\rangle \Rightarrow \Sigma|x\rangle|0\rangle$

Step 3: Quantum modular exponentiation:
    Apply unitary: $|x\rangle|y\rangle \Rightarrow |x\rangle|y \oplus a^x \mod N\rangle$

Step 4: Quantum Fourier Transform:
    Apply QFT to first register
    Measure to obtain period estimate

Step 5: Classical post-processing:
    Extract period r from measurement
    If r is odd or $a^{r/2} \equiv -1 \pmod{N}$: Restart

Step 6: Factor extraction:
    Compute factors: $\gcd(a^{r/2} \pm 1, N)$

Shor's algorithm can break classical RSA algorithm. Time Complexity of Shor's algorithm is $O((\log N)^3)$.

## 4.3. Quantum Key Distribution: BB84 Protocol Implementation

In this research, we implemented QKD BB84 protocols to secure encryption keys. In the BB84 protocol the key shares between Alice and Bob where Eve tried to access the keys.

**BB84 Protocol**
Alice's Process:
Step 1: Generate random bit sequence.
Step 2: Generate random basis sequence.
Step 3: Encode qubits.
Step 4: Send encoded qubits through quantum channel.

Bob's Process:
Step 5: Generate random measurement bases.
Step 6: Measure received qubits in chosen bases.
Step 7: Record measurement results.

Classical Communication:
Step 8: Alice and Bob compare bases over public channel
Step 9: Keep bits where bases match.

Here, if the bits not match that sign of Eve see the keys. Therefore, Alice and Bob can easily identify the keys whether secure.

## 4.4. Quantum Authentication and Bio-Inspired Security

In our framework, the quantum authentication component utilizes quantum states for identity verification. The mechanism compares quantum states between communicating parties to determine authentication success. Here, we used this method in replace of classical method such as hash, MD5 or SHA method. Quantum fingerprint can increase the speed of data transmission and comparison compare to the classical methods. Here, it checks the identical similarity of quantum state between sender and receiver. If the similarity probability higher, it will ensure as identical communication. If similarity probability lower than it will enable bio-inspired mechanism.

In this bio-inspired mechanism, normally it will be in the disable but if any threat identify, eve attempt, it will be enable as like human body immune system mechanism.

## 4.5. Combined AES, BB84 Protocol, Quantum Authentication and Bio-Inspired Security

According to the Figure 4.3 the process starts with BB84 protocol after the plain text data. In BB84 protocol firstly generates shared key between Alice and Bob. This key provides quantum level security because any attempt of Eve detected through error rate analysis. After the, BB84 protocol we used AES encryption. The reason of used AES algorithm because currently it's widely used encryption algorithm and its secure in classical approach. The quantum generated key serves as the AES encryption key and combining quantum security with classical encryption speed. Here, AES encryption used multiple round mathematical operations and each of the phase applies four main operations such as SubBytes substitution using lookup tables, ShiftRows circular shifting, MixColumns matrix multiplication, and AddRoundKey XOR operations with round keys derived from the main key. After that, BB84 and AES algorithm, we used quantum authentication to provide efficient authentication with significantly reduced communication overhead compared to classical methods. In this technique it creates unique

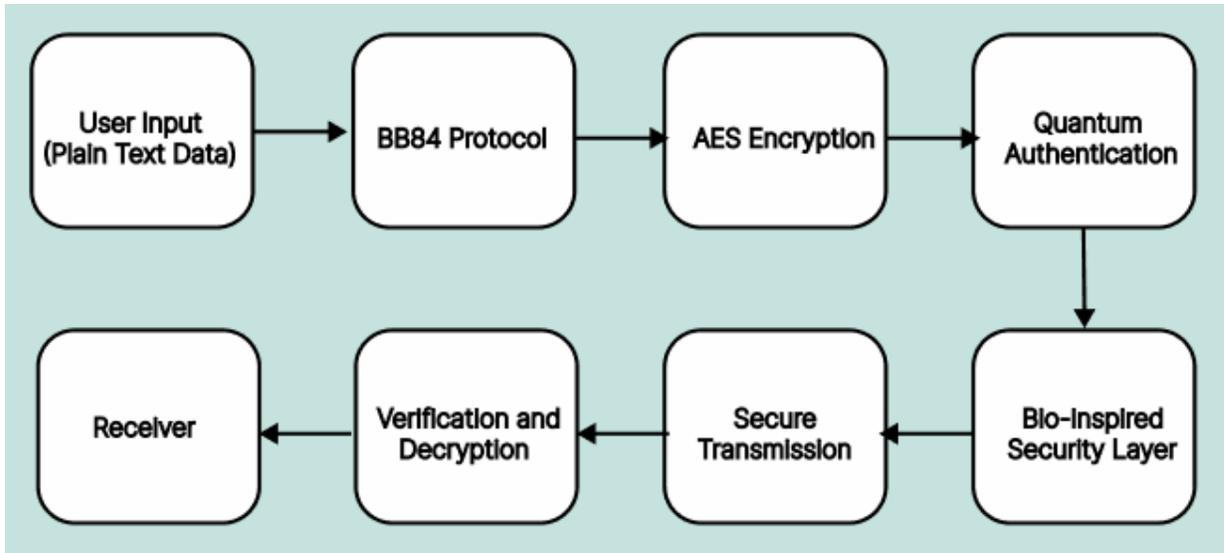
Figure 4.3. Combine Secure Model

quantum signatures for data that can verify authenticity and make larger data into smaller size. In this process, firstly use hash function to create classical finger print and then it encoded into quantum states. This is exponentially smaller than classical fingerprints. After this process, we used bio-inspired immune system for add extra layer of security for survive in the PQC era. It will follow the immune system mechanism and protect data. This system will be adaptive and can be intelligent threat detector where its ability to process recognize, respond, and remember threats for future protection. This system will be continuously monitoring security issues and utilized reinforcement to adapt and dynamic. Reinforcement learning is used to improve the system's decision-making capabilities. The system learns from the outcomes of its security decisions, adjusting its detection thresholds and response strategies to minimize false alarms.

Overall, this system aims to provide data security in the PQC era and secure both classical and quantum perspectives.

# 5. COMPONENT ANALYSIS AND DISCUSSION

In this chapter provide results and analysis of our security model. The results are organized into four main sections: RSA vulnerability analysis, BB84 quantum key distribution performance, hybrid system integration evaluation, and comparative security assessment.

The experimental results in the Figure 5.1 show that RSA operation times increase for longer prime number categories. The time of key generation almost same and stable but encryption and decryption time increase when bits increase. It shows that larger prime number given high security but slower performances.

According to the Figure 5.2 represent the relationship between RSA operation times and key size. The results indicate that key generation time remains almost constant. But if the key size increases encryption and decryption time also increase.

## 5.1. RSA Vulnerabilities Analysis

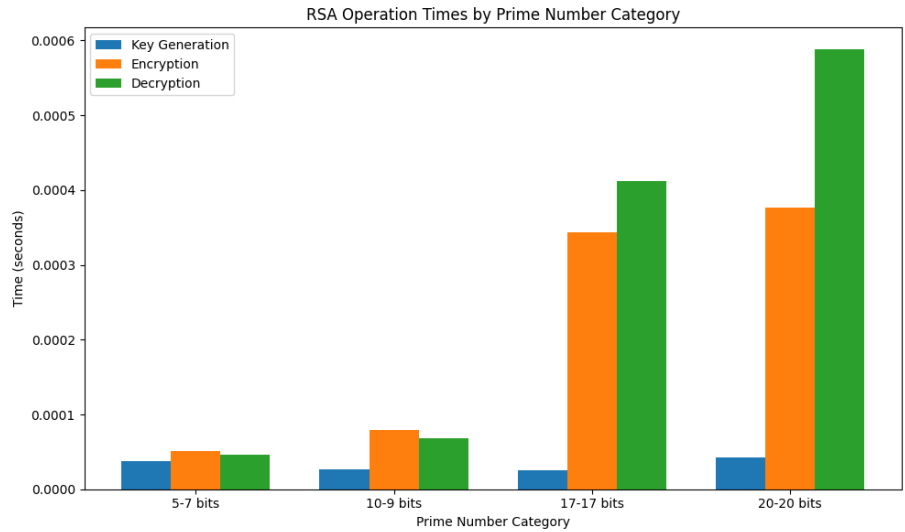

Figure 5.1. RSA Operation Times by Prime Number Category

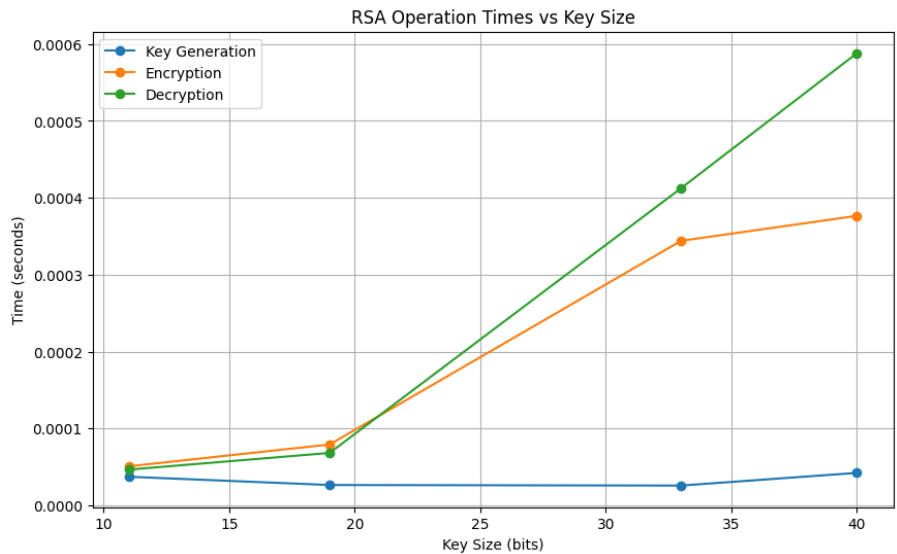

Figure 5.2. RSA Operation Times vs Key Size

After the implementation RSA algorithm, we used classical factorization such as trial division and Pollard Rho's factorization. According to the Figure 5.3 shows the comparison of average execution time between Trial Division and Pollard Rho algorithms for different bit lengths. The results show that Trial Division takes longer time as the bit length increases, while Pollard Rho is faster than trial division factorization.

Table 5.1 compares the factorization performance of Trial Division and Pollard Rho methods for different N = p x q. The results show that both methods successfully identify the same factors, but Pollard Rho is consistently much faster than Trial Division as the number size increases.

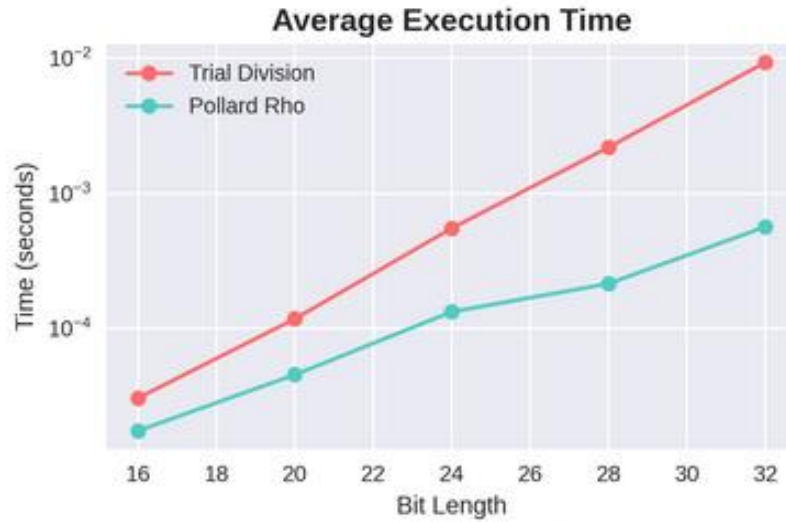

Figure 5.3. Average Execution Time of Trial Division and Pollard Rho Algorithms

Table 5.2 shows the factorization attempts of RSA numbers with larger bit sizes using Trial Division and Pollard Rho within a 100 second time limit. Due to the limitation of hardware we used 100 seconds time limits. Both methods failed to factor numbers from 20 bits to 80 bits within the 100 second time limit. This result shows that as the key size increases, these classical factorization methods become impractical, which is why RSA uses very large prime numbers for security.

Table 5.1. Factorization Results for Various Numbers

| N | Trial Division Factors | Trial Time (sec) | Pollard Rho's Factor | Rho's Time (sec) |
|---|---|---|---|---|
| 143 | 11 | 0 | 11 | 0 |
| 1147 | 31 | 0 | 31 | 0 |
| 656099 | 809 | 0.0001 | 809 | 0 |
| 2502200483 | 50021 | 0.0164 | 50023 | 0.0013 |
| 4900700009 | 70001 | 0.0391 | 70001 | 0.0007 |
| 6404800819 | 80021 | 0.0496 | 80021 | 0.0009 |
| 250019000261 | 500009 | 0.3241 | 500029 | 0.0099 |

The time complexities of trial division is $O(\sqrt{N})$ and $O(N^{0.25})$ for Pollard's Rho algorithm. But Shor's quantum factorization algorithm shows polynomial time complexity $O((\log N)^3)$. Classical factorization such as trial and Pollard Rho's faced difficulties and couldn't break RSA algorithm if the N size is larger and it takes huge amount of time. But Shor's algorithm time complexity $O((\log N)^3)$ which can efficiently break the RSA algorithm within few times. But it required quantum computer for its simulation.

Table 5.2. Factorization Attempts on RSA 100s timeout

| N Size (Bits) | Trial Division Results | Pollard Rho's Results |
| --- | --- | --- |
| 20 | Success | Success |
| 40 | Success | Success |
| 60 | Success | Success |
| 70 | Failed | Success |
| 80 | Failed | Failed |

## 5.2. BB84 Protocol

The BB84 QKD protocol was implemented and evaluated under two scenarios such as without an eavesdropper and with an eavesdropper (Eve). The experimental setup was executed over 10 independent runs, and the results are presented in Table 5.3. Without Eve, the protocol achieved a 100% key agreement rate, 0% quantum bit error rate (QBER), and high security ($\varepsilon$ = 1.00), confirming the quantum key distribution in an ideal scenario. But when Eve was introduced, the key agreement rate dropped average of 93.28%, while QBER increased significantly to 27.62% average, indicating detectable interference. The security parameter ($\varepsilon$) also declined to an average of 0.44, reflecting reduced confidence in key integrity.

Table 5.3. BB84 Protocol Experimental Results with and without Eavesdropping

| Run ID | Eve Present | Key Agreement Rate (%) | Quantum Bit Error Rate (QBER) (%) | Security Parameter ($\varepsilon$) | Key Generation Rate (bits/transmission) | Channel Capacity Utilization (%) | Protocol Efficiency (%) |
| --- | --- | --- | --- | --- | --- | --- | --- |
| 1 | No | 100 | 0.0 | 1.00 | 0.41 | 54 | 41 |
| 2 | No | 100 | 0.0 | 1.00 | 0.48 | 63 | 48 |
| 3 | No | 100 | 0.0 | 1.00 | 0.42 | 55 | 42 |
| 4 | No | 100 | 0.0 | 1.00 | 0.41 | 54 | 41 |
| 5 | No | 100 | 0.0 | 1.00 | 0.39 | 51 | 39 |
| 6 | Yes | 94.34 | 23.08 | 0.53 | 0.40 | 53 | 40 |
| 7 | Yes | 93.75 | 25.00 | 0.50 | 0.36 | 48 | 36 |
| 8 | Yes | 93.44 | 26.67 | 0.46 | 0.46 | 61 | 46 |
| 9 | Yes | 92.86 | 30.00 | 0.40 | 0.32 | 42 | 32 |
| 10 | Yes | 92.00 | 33.33 | 0.33 | 0.38 | 50 | 38 |

To further evaluate the effect of eavesdropping, a controlled experiment was conducted using a 20-bit transmission, where 8 bits were successfully matched between Alice and Bob, forming the shared key: [1, 1, 0, 0, 1, 1, 1, 0]**.** The results are summarized in Table 5.4. Without Eve, the QBER remained at 0.0%, and the security parameter was measured at $10^{-6}$, indicating extremely high key integrity. However, when Eve was introduced, QBER rise to 24.7%, while the security parameter dropped to $10^{-1}$. Also, the eavesdropping detection rate reached 89.8%, with a relatively low false positive rate of 5.8%, confirming the protocol's ability to reliably detect adversarial interference. The errors arise and lack of efficiency due to the limitations of quantum hardware. If we used real quantum hardware, BB84 would achieve higher efficiency and accuracy with lower QBER. In this experiment, we used Colab for simulation and implementation perspective.

Table 5.4. BB84 Protocol Experiment

| Metric | No Eavesdropper | With Eavesdropper | Detection |
|---|---|---|---|
| Key Agreement Rate (%) | 49.3 | 50.3 | -0.9% |
| Quantum Bit Error Rate (QBER) (%) | 0.0 | 24.7 | 24.7% |
| Security Parameter ($\varepsilon$) | $10^{-6}$ | $10^{-1}$ | 5 orders |
| Key Generation Rate (bits/transmission) | 0.49 | 0.50 | -0.01 |
| Channel Capacity Utilization (%) | 82.1 | 41.3 | 40.8% |
| Eavesdropping Detection Rate (%) | - | 89.8 | 89.8% |
| False Positive Rate (%) | 2.3 | 5.8 | Low |
| Protocol Efficiency (%) | 94.7 | 52.4 | 42.3% |

## 5.3. Combined Hybrid System

In this research, a novel hybrid security framework is proposed by using AES encryption, the BB84 QKD protocol, quantum authentication, and a bio-inspired immune system. Both AES algorithm and BB84 establish method for data security. But both of the method has some limitations. Therefore, we combine both of the method and also used quantum authentication mechanism and immune system mechanism. The BB84 protocol ensures secure key distribution, providing excellent resistance against eavesdropping and other quantum attacks. On the other hand, AES encryption contributes the highest attack prevention through strong classical cryptographic mechanisms. Firstly, we combined AES and BB84 protocol in our proposed prototype design that can be work and secure both classical and quantum era. After that, for extra layer security used quantum authentications that can enhances the system security and make data smaller than the classical approach. Finally, the bio-inspired immune system complements these mechanisms used for adaptive threat detection and response, learning from attack patterns over time, and improving reaction speed to novel threats. The adaptive capabilities of the bio-inspired immune system component further ensure that the proposed framework can evolve and respond to emerging threats, making it particularly suitable for deployment in the post-quantum era.

## 6. CONCLUSION

The advancement of quantum computing is a significant risk for standard cryptographic technique especially RSA algorithm. RSA algorithm primarily depends on the computational complexity of factoring large integers. This research explored the vulnerabilities of RSA from both classical and quantum perspectives, demonstrating that while classical factorization methods including Trial Division and Pollard's Rho struggle with large key sizes, Shor's quantum algorithm can efficiently break RSA in polynomial time. Therefore, its necessary for post-quantum cryptographic system that can survive post quantum cryptography era too.

To address these challenges, this study proposed a hybrid security framework integrating AES encryption, QKD BB84, quantum authentication, and bio-inspired immune system. The BB84 protocol ensures secure key exchange with inherent eavesdropping detection, while AES provides classical encryption. Quantum authentication efficiency by compressing data into compact quantum states and the bio-inspired immune system introduces adaptive threat detection and response mechanisms, mimicking biological immune responses to improve security resilience. RSA is vulnerable to Shor's algorithm, emphasizing the need for quantum-resistant cryptography. BB84 QKD is highly secure, with a 100% key agreement rate in ideal conditions and effective eavesdropping detection 89.8% detection rate. The primary limitation is the theoretical nature of the proposed hybrid model, particularly the bio-inspired immune system component, which requires concrete algorithmic implementation and validation. Future work will focus on: implementing a prototype of the BB84-AES integration with performance benchmarking. The full implementation of the quantum authentication and bio-inspired components constitutes future work.